\documentclass
{birkmult}
\usepackage[latin1]{inputenc}
\usepackage{textcomp,graphicx,color,hyperref,afterpage}
\def\Newitem{\par\textemdash\kern.3em\ignorespaces}

\begin{document}
\fboxsep1pt
\newdimen\zegoodwidth
\zegoodwidth=\columnwidth
\advance\zegoodwidth-2\fboxrule
\advance\zegoodwidth-2\fboxsep
\title[Digital Mathematics Libraries: The Good, the Bad, the Ugly]
 {Digital Mathematics Libraries:\\ The Good, the Bad, the Ugly}
\author
{Thierry Bouche}

\address{%
Université de Grenoble~I \& CNRS,\\
Institut Fourier (UMR 5582) \& Cellule MathDoc (UMS 5638),\\
BP 74,  38402 St-Martin-d'Hères Cedex, France}

\email{thierry.bouche@ujf-grenoble.fr}

\urladdr{\url{http://www-fourier.ujf-grenoble.fr/~bouche/}}
\subjclass{00A99}


\date{January 25, 2009}

\begin{abstract}
  The idea of a World digital mathematics library (DML) has been
  around since the turn of the 21th century. We feel that it is time
  to make it a reality, starting in a modest way from successful
  bricks that have already been built, but with an ambitious goal in
  mind. 

  After a brief historical overview of publishing mathematics, an
  estimate of the size and a characterisation of the bulk of documents
  to be included in the DML, we turn to proposing a model for a
  Reference Digital Mathematics Library---a network of institutions
  where the digital documents would be physically archived. This
  pattern based rather on the bottom-up strategy seems to be more
  practicable and consistent with the digital nature of the DML. After
  describing the model we summarise what can and should be done in
  order to accomplish the vision.

  The current state of some of the local libraries that could
  contribute to the global views are described with more details.
\end{abstract}

\maketitle

\section{The Mathematical Literature}

\begin{flushright}
  \begin{minipage}{.8\linewidth}\footnotesize
    Les mathématiciens se contentent de mettre leur production à la
    disposition de tous, comme sur des étagères où l'on peut venir se
    servir.\footnotemark

    \hspace*{\stretch{1}}Jean-Pierre Serre (according to Michel
    Broué~\cite{broue}).
  \end{minipage}
\end{flushright}

\footnotetext{``Mathematicians just make their results freely available, as if
    they were on shelves where anyone can fetch them.''}

\subsection{Stakes}

Mathematics is unique among the hard sciences in its dependence on its
scholarly literature.  Mathematicians and users of mathematics rely
crucially on long-lasting access to original validated research
articles, monographs and textbooks. The mathematical corpus, however,
is more than just a collection of works; it is a complex network of
interconnected items, some centuries old but still valid and relevant,
each referring or related to, dependent upon or supporting each other.

The knowledge society needs reliable foundations, which implies that
published mathematical results be checked, and that the checked
versions be stored indefinitely. The storage must be carefully
organised, with a clean and detailed catalogue, so that any one of
those items can be referred to at any time later on, with no
ambiguity. A dependence tree should be constructed as well, so that
new material partly based on old one still can be trusted.

Because users of mathematics do not necessarily rely on the
{current} mathematical output, it should also be easily
accessible over long periods of time. 

Fads and trends go: the criteria for eligibility in such an archive
should not be the popularity of an author or a subject at its time,
but the conformance to rigorous standards of production and
validation. Each new result with an original proof that has been
carefully checked by independent experts can become a crucial
reference for unexpected developments, and find tremendous
applications in other scientific as well as technological areas. A
number of algorithms, arithmetic theorems, effective results had been
studied before the first computers were even conceived. These schools
were sometimes considered foolish or exploring dead-ends. But the
rapid development of computer science or cryptography could not
have been achieved if this theoretical background had been lost.

\subsection{The reference library}
\label{sec:reflib}

These facts put together explain why mathematicians have always taken
great care for their libraries, which are the central infrastructure
of all math labs worldwide\footnote{%
  A recent study over French mathematical laboratories concluded that
  about one third of their expenditures went into their library's
  budget \cite{dalbo-etal}.}%
.  The ideal library should be exhaustive, acquire promptly new
publications, and enjoy wide opening hours and low administrative
barriers to occasional visitors from other locations or disciplines.
Thanks to the stubbornness of the mathematical community, those
(paper) libraries approximating fairly the ideal situation are not
few, and almost evenly distributed in the developed countries.
However, each laboratory has idiosyncratic bias towards some subjects,
and a limited budget, so that no-one among these lab libraries
provides a full reference to the mathematical corpus. Luckily though,
interlibrary loan assembles these dispersed libraries into one virtual
global resource providing more or less the expected feature, as long as
a precise union catalogue makes it possible to locate the items
somewhere in the network.

It should be stressed here that the value of this reference library
system does not only reside in the ability for researchers to have a
fast access to the resources that they most need in their daily work.
On the contrary, mathematical original research has a very small
audience, and is seldom consulted, but basic sciences could not be
performed without the reliable foundations provided by the
mathematical corpus as a whole.

\subsection{The digital era}

The birth of electronic communication at the end of 20th century,
which has become an ubiquitous, almost exclusive mean of disseminating
knowledge nowadays, did not change dramatically science's needs. It
has opened new opportunities for easier, faster dissemination, and
more powerful discovery of scientific results.

Most of the traditional tools that working mathematicians and
scientists have been using for years enjoy an electronic counterpart:
informal discussions with colleagues can be handled through emails and
blogs, in addition to face-to-face meetings; more formal preprints are
disseminated through eprints archives. Formal publication in a
refereed journal is still the mainstream for getting an independent
evaluation of new results and their proofs: this provides the original
articles that will enable reference for further research. The
reviewing journals \emph{Zentralblatt} and \emph{Mathematical
  Reviews}, which have long been the main gateways to the recent
literature, thanks to independent reviews and subject classification,
have been converted to databases that can now be searched over their
whole history, and often provide links to the actual digital resources
they index: Zentralblatt MATH (ZM \cite{zmath}), MathSciNet (MSN \cite{msn}).

For  researchers, it is routine to go out hunting for one of
those references that is needed to understand a proof which settles on
its results, or that has been brought to one's attention by a
colleague, a hit in some search engine, etc. This is something that
always needed some expertise, and is made easier by a well-organised
library, and the help of librarians. Many of these references exist
today in digital format, thanks to retrodigitisation efforts and the
generalisation of electronic publishing. It is not an easy task,
however, to determine whether one given reference is available
digitally or not, where to search for it, how to locate it, which of
the sometimes numerous digital versions is the one that was precisely
referred to (and what differences really exist between the versions).
Finally, once a resource has been painfully located, one faces
frequently the deception that no access is allowed (which will quite
often be an artifact of the path followed to discover it!).

\section{The Mathematical Corpus}

We call \emph{mathematical corpus} the set of all published
mathematical texts (possibly endowed with an oriented graph structure)
in the sequel.

\subsection{Chronology}

Although it is commonplace to start the history of mathematics in
Mesopotamia nearly 4000 years ago, the tradition on which the current
corpus of written mathematics settles started with Euclid's
\emph{Elements} 2500 years ago. Manuscripts and copies are the media
of the time till Gutenberg. Scientific communication is dominated by
book publishing and personal correspondence until the end of the 17th
century, when scholarly journals are invented in France and Great
Britain. Specialised journals appear some time later (1810 sees the
publication of the first mathematics-only journal in N\^\i mes:
\emph{Annales de math\'ematiques pures et appliqu\'ees}, edited by
Joseph Gergonne). This model then spreads across Europe, where
language barriers are still high: an important effort is devoted to
translating or abstracting results from foreign journals up to early
20th century. 

At the end of 19th century, mathematicians start to feel that their
discipline has grown so much that no-one can keep track of the
published discoveries. Many enterprises are started in order to build
tools allowing the working mathematician to find his path in the
literature. Librarians write reference catalogues in Berlin or London,
the \emph{Jahrbuch \"uber die Fortschritte der Mathematik}, first
reviewing-only journal, appears in Berlin in 1868, the French
mathematical society launches the \emph{R\'epertoire bibliographique
  des sciences math\'ematiques}, which is a list of articles published
during the 19th century. To help users on their way, classification
schemes are developed.

During the 20th century, mathematics keep growing while the world is
getting smaller: journals tend to be international in audience and
authorship. At the end of the period, it is considered that about
100,000 new references are published a year, through a core of some
600 math-only journals, and a myriad of other channels (proceedings,
books, scientific serials with a wider coverage than math-only\ldots). 

Professional desktop publishing is introduced in the mathematical
field in the 1980s, with Donald Knuth's \TeX. As a consequence,
instant unmediated dissemination of mathematical writings becomes
possible, while the serial pricing crisis puts some pressure on the
libraries' subscriptions.  It becomes also clear that all aspects of
scientific information is getting digital: After typesetting,
prepress, catalogues and reviewing journals are turned into databases.
Electronic publishing becomes ubiquitous at the end of the century,
and digitisation projects try to bring back the already released
material into the new paradigm.

At the beginning of the 21th century, the entropy is still growing. It
turns out that the mathematical literature is quite often available in
a dual format: 
\Newitem printed on paper, sometimes from digital source (like
print-on-demand);
\Newitem in digital format, sometimes as scanned images from paper.

These two sets overlap a lot, and more so every day, but it is
doubtful they will ever converge. On one hand, although massive
digitisation has been performed, many isolated items will be left
apart. On the other hand, it is likely that more and more
electronic-only items will be published.

\subsection{Size}
\label{sec:dmlsize}

It is estimated that 2.5 million items belonged to the mathematical
corpus when last century ended, and that 100,000 new items have
appeared each year since then.

Less than a fifth of the total was published before the 20th century,
and more than a half after 1950, which means that current publishing
model covers quite a big portion of the lot. 

The vast majority (around 80\;\% of 20th century output) of those
items are journal articles. Conference proceedings and collective
books amount for another 10\;\%. From a relatively large corpus of
such documents, we can infer that the mean number of pages for
articles is 20, while it should be above 100 for books. Assuming thus
30 pages per item, we get an estimate of 3~million items today,
spanning 100~million pages for the whole corpus, with a current annual
growth of 100,000~items over 3~million pages.

\smallskip

As regards the portion of the mathematical corpus which exists in
digital format, we focus on the subset of items held either by their
publishers or academic digital libraries. This excludes personal
collected works at author's Web pages, as well as large reservoirs
such as Google Books \cite{gglprint}, where the digital files are
metadata to some original rather than reliable primary sources. We
estimate the existing content to hold around 1,5~million items,
covering about 15~million pages. The discrepancy between the average
page count of digital items compared to the whole corpus can be
explained by many factors.
\Newitem Journal articles have been the core of many digitisation
  projects, and are those items that are natively produced digitally
  since 1997 while books are way behind in the digitisation process.
\Newitem Some publishers and digitisation projects register every
  information bit (book review, letter to the editor, back
  matter\ldots{}) as a mathematical article if it is excerpted from a
  journal belonging to one of their math package, which is not the
  criterion used by the traditional catalogues and reviewing journals
  that were used to estimate the overall size of the legacy corpus.
  The reviewing databases are for once almost in agreement on the
  number of those mathematical items they register which have a DOI,
  as they both have around 1~million such DOIs. This is somewhat less
  than the figures advertised by content providers which are known to
  use DOIs for all their items (mostly publishers in our context, plus
  JSTOR and project Euclid). The number of old articles absent from
  the databases is too low to explain the discrepancy, so we can infer
  that content providers tend to have a much more relaxed definition
  of mathematical items than the one of the reviewing databases and
  the mathematical community in general.

 Nevertheless, one can estimate that the digital corpus already
 amounts to somewhere between one sixth and third of the whole, which
 is considerable!

\subsection{Geographical and linguistic span}

From middle age to 19th century, Europe is the centre of natural
sciences. The mathematical tradition started in Greece and India has
come back through Arabic scholars and this is where the foundations of
modern science will be shaped.  This leaves us with numerous written
records: manuscripts, books, private letters, transactions, serials.
While Latin has been the \emph{lingua franca} of all scholarly
writings during this period, vernacular idioms come soon
into the picture, then structure themselves as national, regional or
international depending on various factors.

The core mathematical knowledge has been produced and stored in
Europe, spread across many countries and languages. It became truly
international at the end of the 19th century, which is exemplified by
the birth dates of the national mathematical societies (Bohemia: 1862,
United Kingdom: 1865, France: 1872, USA: 1888, Germany: 1890, etc.).
The first International Congress of Mathematicians was held in
Z\"urich in 1897, with 197 members from 15 European countries plus 7
members from the USA. The International Mathematical Union was formed
in 1920. Up to the 1980s, virtually any mathematical journal would
accept, in addition to the local language, a paper written in English,
German, Italian, or French. International journals still published
German and French articles by the nineties; it seems that French is
currently the only (rare) alternative to English in places like
\emph{Annals of Mathematics} or \emph{Inventiones Mathematic\ae}.
However, the existing mathematical corpus is not reduceable to a
single, or even a handful number of languages.

\section{A proposed model for a Reference Digital Mathematics Library}
\label{sec:dml}
\begin{flushright}
  \begin{minipage}{.8\linewidth}\footnotesize
    ``In light of mathematicians' reliance on their discipline's rich
    published heritage and the key role of mathematics in enabling
    other scientific disciplines, the Digital Mathematics Library
    strives to make the entirety of past mathematics scholarship
    available online, at reasonable cost, in the form of an
    authoritative and enduring digital collection, developed and
    curated by a network of institutions.''  

    \hspace*{\stretch{1}} DML project vision, Cornell library, 2002.
  \end{minipage}
\end{flushright}

\subsection{The Vision}

Taking into account the needs of the mathematicians, and of science at
large, as summarised in \S~\ref{sec:reflib}, and the fact that the
paper library is slowly declining into a dead archive, we infer that
there is a need for a new infrastructure providing the facility of the
reference mathematical library in the digital paradigm, which will be
called RDML in the sequel.

When we refer to a digital library, we mostly refer to a traditionally
organised library with digital objects on its shelves.  This means
that the stress is on the traditional library functions, rather than
on fancy digital stuff. The main outcome of the envisioned library
service would be to set-up a network of institutions where the digital
items would be physically archived.  Each institution would provide
its own contribution to the network through various interoperability
devices (some socially oriented, like training or policy making; some
technically oriented, like metadata harvesting, cross-repositories
linking\ldots).  Each institution would take care of selecting,
acquiring, developing, maintaining, cataloguing and indexing,
preserving its own collections according to clear policies: it should
have the role of a reference memory institution for a well defined
part of the mathematical corpus. It would provide and control access
to the full texts, when needed.  The \emph{physical} operations on
collections' objects (acquisition and delivery) would be \emph{local},
and entirely performed at the relevant institution. At the local
level, we would not consider virtual libraries referencing third party
objects with no control over it. The network of institutions would
make it possible to assemble a \emph{global}, \emph{virtual} library
providing a one-stop gateway to the distributed content through
user-friendly retrieval interfaces. 

That would fit well with the vision of a central (cyber)infrastructure
of the global networked mathematics department, where the meeting room
with a chalk board is one of the scientific cafés that emerged in
Web~2.0, and the library has its collections ready for direct
references when needed in the discussion.

\def\mDML{RDML}
\subsection{The Design of the RDML}

The implementation of the RDML vision will necessarily be an
incremental process. The first step will be successfully completed
when two local institutions start sharing enough metadata in order to
be searchable in a single database.

The \mDML\ network would thus constitute a distributed digital
repository of validated mathematical original research texts from many
sources. The content gathered would be either retrodigitised from
legacy paper publications, or born-digital contributed by its
publisher to one of the \mDML\ local institutions. Let us recall that
the main objective is to recreate most of the traditional functions of
a legacy mathematics department library in the digital paradigm, while
taking advantage of the format to set up unique services that would
address the specific issues faced in the management of a heavily
multilingual mathematical knowledge.  The main service to the
community would be the ability to discover easily, enjoy seamless
access, and refer to a given text permanently.  These services would
be tailored for the end user (i.e.  researchers), but also have
automated counterparts in order to be interoperable with the other
important research infrastructures (like reviewing databases,
publisher's websites, institutional repositories\ldots): it would be a
major component of the emerging eScience paradigm where mathematical
scholarship is needed.

\subsection{Institution selection}

The public presence of the \mDML{} would be built on top of a unique
database registering objects in the contributed \emph{physical}
digital libraries, each of these being hosted at one of the
participating institutions.  An institution should be a scientifically
reliable, long-standing, not-for-profit organisation with a clear
policy about long-term archiving and access.  It would be the
responsibility of each of these institutions to negotiate the licenses
allowing them to work with the content they care for: archiving,
indexing, possibly migrating formats and upgrading metadata, providing
eventual open access.

A consequence of this policy is that a considerable part of the
existing digital mathematical content could not be registered in the
\mDML\ right now. This is a concern that should be addressed later on
when a widely agreed upon policy is formalised, and a critical mass is
attained that conforms to this policy.

\subsection{Content selection}
\label{sec:cs}

The RDML vision is to revive the concept of a mathematics department's
library in the digital realm, aiming at comprehensiveness, but
avoiding redundancy to the extent possible, thanks to a network of
collaborating centres. The main criteria for eligibility should be
easy to establish: validated mathematical texts form the core of the
collections. They range from books, Ph. D. theses, refereed journal
articles, to seminar or conference proceedings. As we are talking
about a very basic resource, it would not hurt if the collections
happen to cover a wider field than core mathematics. What could be
unfortunate would be to have duplicates from various partners that do
not match identically, which implies a strong metadata policy in order
to distinguish editions of books, e.g. The main point here is that the
RDML is concerned with scientifically validated material having passed
a publishing process with some sort of quality insurance. Volatile
material that is not meant to be relevant after a short while would be
on a low priority. 

On the technical side, the master files to be stored should be in open
document formats with no restrictions so that the content can be used
over the long term, whatever processing on the files could be needed.

Copyright and licences have also consequences over long term use and
accessibility of the scientific texts. They will have to be carefully
considered for each collection, sometimes for each item, to help decide
whether they make acquisition worthless.

\subsection{Content acquisition}

At our local digital libraries, acquisition means to ingest computer
files into their information system. There is a large variety of
sources for these files: they can be produced from paper by
digitisation projects, or entirely produced by some external entity.
The acquisition process mostly consists in standardising formats from
this variety of inputs. A minimal item is a full text with some
metadata associated to it. 

As no electronic format has still emerged that permits to capture the
whole meaning of a real-world mathematical text, the above mentioned
full text will indeed be stored as some sort of graphic-oriented file
format, typically a PDF or DjVu file.

In many cases, extra work is needed in order to generate all the files
and formats needed for local operation, as well as to restructure or
extend provided metadata.

\subsection{Metadata}

In order to fulfil mathematician's most basic needs, a minimum
metadata set should be defined for every archived item. Of course, the
typical elements in a library catalogue should be captured, such as
author names, title, and full bibliographic reference when
applicable. In any case, as we are dealing with published material, it
is very important for any subsequent use that the publication vector
be identified, as well as the original date of publication. 

It is often the case that many items are published by the same author
under the same title through different channels. Not mentioning
preprints, you can encounter a short announcement, a seminar talk or a
full length detailed paper. The status of the archived text must be
clear from the available metadata, in view of its expected use by
scholars.

Abstracts, key words, mathematical subject classification, links to
related resources---the most prominent being probably the cited
references---would enable considerably the further interoperability.

A purely textual version of the ``full text'' is generally considered
a metadata as well. In this case, the borderline between data and
metadata is not obvious, depending how much differ the actual
mathematical meaning beared by the graphical version of the full text
and the textual one derived from it (by OCR or text extraction, e.g.).

\subsection{Interoperability}

For the collections to enjoy wide visibility and really serve their
purpose as a reference facility, they must be integrated together, and
interoperable with the professional tools such as the reviewing
databases. They should also be ready for future infrastructures that
could set up different retrieval mechanisms than those foreseen today.
This requires metadata standards and policies for sharing them. 

\subsection{Access options}

As concerns the access to the collections, an obvious difference
between a paper and a web-based digital library is the geographical
constraints put on their patrons. If an academic library gets a paper
subscription to a journal, and provides free access to its patrons,
the company that sells it does not expect to go out of business. In
the digital world, if the library serves on the web all the articles
it gets legally through its subscription, it could be the last one the
publisher ever sells\ldots 

This extreme example shows that we have to find some path between an
economic model where every commercially published scholar's work is
privately-owned for ever, and an open library free to anyone.

My observation is that a publisher's mission is to invest in creating
quality new content while long term curation is rather left to public
bodies. Even backfile digitisation resulted in a new product for which
a market was apparently waiting; the maintenance of such services over
the long term is problematic and could prove too costly once the
return on the initial investment is obtained and the fad has gone.

The proposed policy to fit the expectations of copyright holders and the
scientific community is to grant eventual open access to anyone to the
mathematical corpus. A suggested implementation is to define, for each
item in the corpus, a moving wall preventing access until some delay
has expired after its publication.

The moving wall time lag needs not to be uniform over all the item types,
it will in any case certainly need to be adjusted over time. It is an
effective way to make a substantial portion of the mathematical
heritage freely accessible to anyone, while leaving a lot of room for
business on the other end of the time line.

\section{Challenges}

\subsection{Selection}

A high proportion of the mathematical corpus is already available in
digital form. The AMS Digital Mathematics Registry \cite{dmr} lists
1938 journals from 392 sources that have at least part of their
articles available digitally. Ulf Rehmann's registry \cite{rehmandml}
counts 297 digitised serials and more than 4500 digitised books
summing up more than 5~million pages. One can estimate the digitised
mathematical corpus to span over 10~million pages, most of it
privately owned.  Every item authored in the 21th century can be
suspected to have a digital source. For recent material, the selection
criteria of the reviewing databases seem to satisfy the community. For
older material, many catalogues exist.

The ERAM project \cite{eram} in Germany had a very interesting
approach, as it consisted in the digitisation of the \emph{Jahrbuch
  über die Fortschritte der Mathematik} into a database, as well as
the core journals indexed there. Thus the selection criterion also
provided the metadata. The RusDML \cite{rusdml} project followed the
same pattern.

The idea of selecting some sort of ``cream of mathematics,'' that would
represent 5\;\% or so of the whole corpus, is simply wrong.

\subsection{Acquisition}

At the early days of electronic edition, stakeholders  believed
that no content would ever be free anymore, like it had been the case
for centuries with paper copies hold in academic libraries where no
patron needed a valid license or a fresh subscription to access the
volumes and read their inspiring content.  It seems that this
tentative has succeeded to the point where one can read in
authoritative studies apparently objective statements like the
following one: ``In the print era, libraries were acquiring print
journals and took in charge their preservation so that they remain
accessible to their user community in the long term. In the digital
era, libraries and their user community are licensed online access to
electronic journals for a determined and limited duration.''
\cite{ecpubstudy}

This situation is very unsatisfactory and dangerous for the long term
preservation and access to the research published today. Moreover,
backfile digitisation performed by commercial entities, which end up
in packages that are marketed by those entities, might create
``retro-privatisation'' through new rights gained over collections
while they did not necessarily own any rights over the old paper
versions.  For instance, access to a text that is in the public domain
may become illegal to non-subscribers because the file that bears it is
newly copyrighted, or a publisher that just acquired a long lasting
independent first class journal makes all its intellectual heritage
its property at once, when adding it to its online offer.  When even
very old texts become unavailable unless you have a specific
subscription for each of them, this places the whole system of
referencing and linking at risk. This places also a high burden on
scholars from everywhere in the world to achieve their task.

This is why digital mathematics need a simple and reliable archiving
system which is not aimed at profit, but at sustainability. This would
be achieved by the \mDML\ network of partners, acting like memory
institutions, each one committed to acquire and curate a local subset
of the mathematical corpus. It does not seem necessary to endow the
above expression with a too precise definition. It suffices to
acknowledge how the current DML efforts have structured themselves
spontaneously in this respect: national borders having an important
impact on funding, languages, human proximity, many nation-wide
projects have emerged.  Many of those national projects deal with
``foreign'' content (content is typically internally deeply
international anyway, as current mathematical research crosses
boundaries) but, we have to reckon that national forces are still
rather active (G\"ottingen's GDZ \cite{gdz} digitised Swiss and Czech
journals, which have then been shared with the ``national'' DML
projects---SEALS \cite{seals} and DML-CZ \cite{dmlcz},
respectively---so that they can be bundled with the other sources from
the same origin, or upgraded with newly published articles). On the
other hand, other kinds of local projects are in existence, like:
subject oriented (algebraic geometry, e.g., which has always been at
the leading edge of the move to electronic literature), or author
oriented (electronic collected works, e.g.).

\subsection{Metadata quantity and quality}

\begin{figure}[tp]
  \centering
  \fbox{\includegraphics[width=\zegoodwidth]{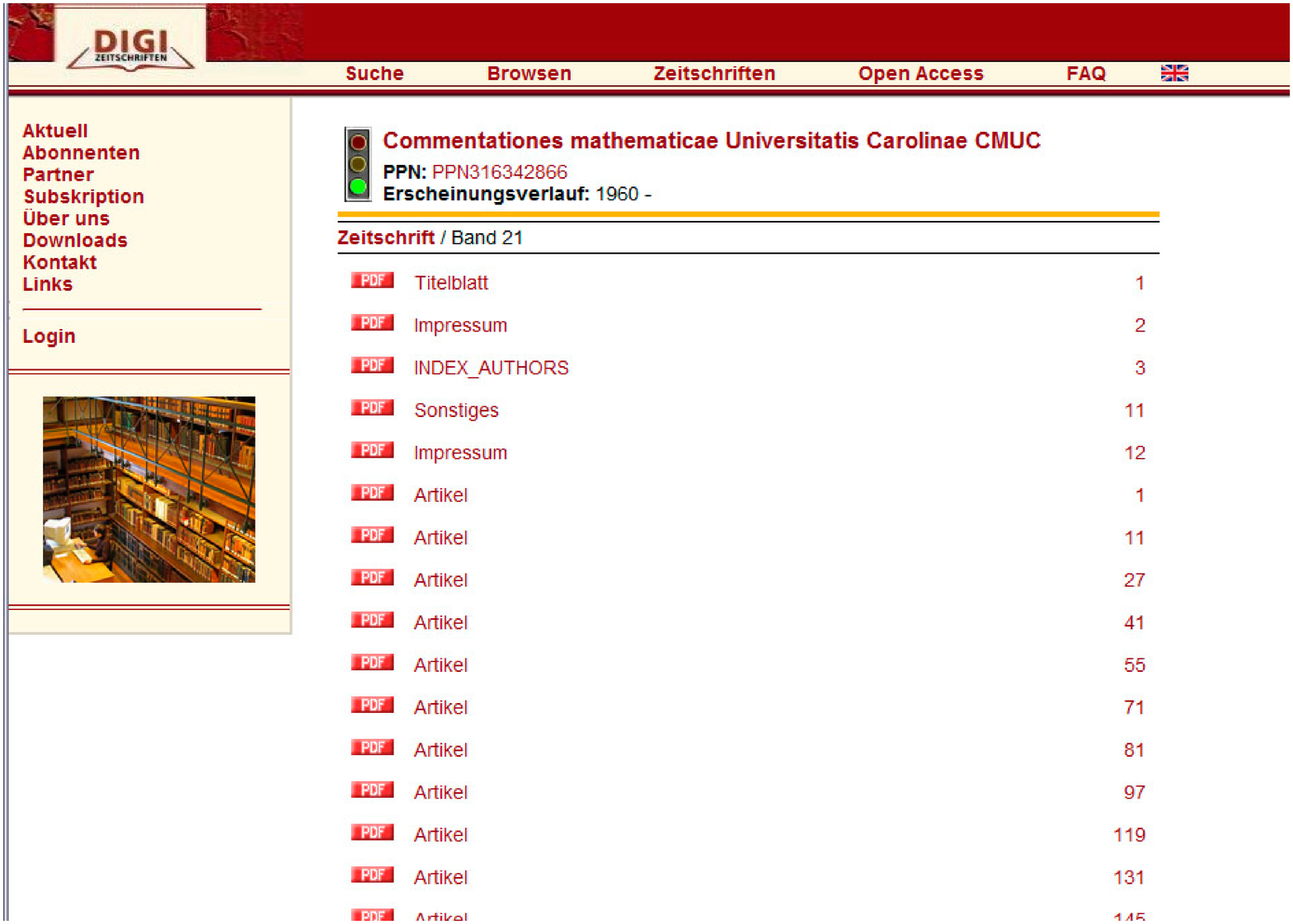}}\\[3mm]
  \fbox{\includegraphics[width=\zegoodwidth]{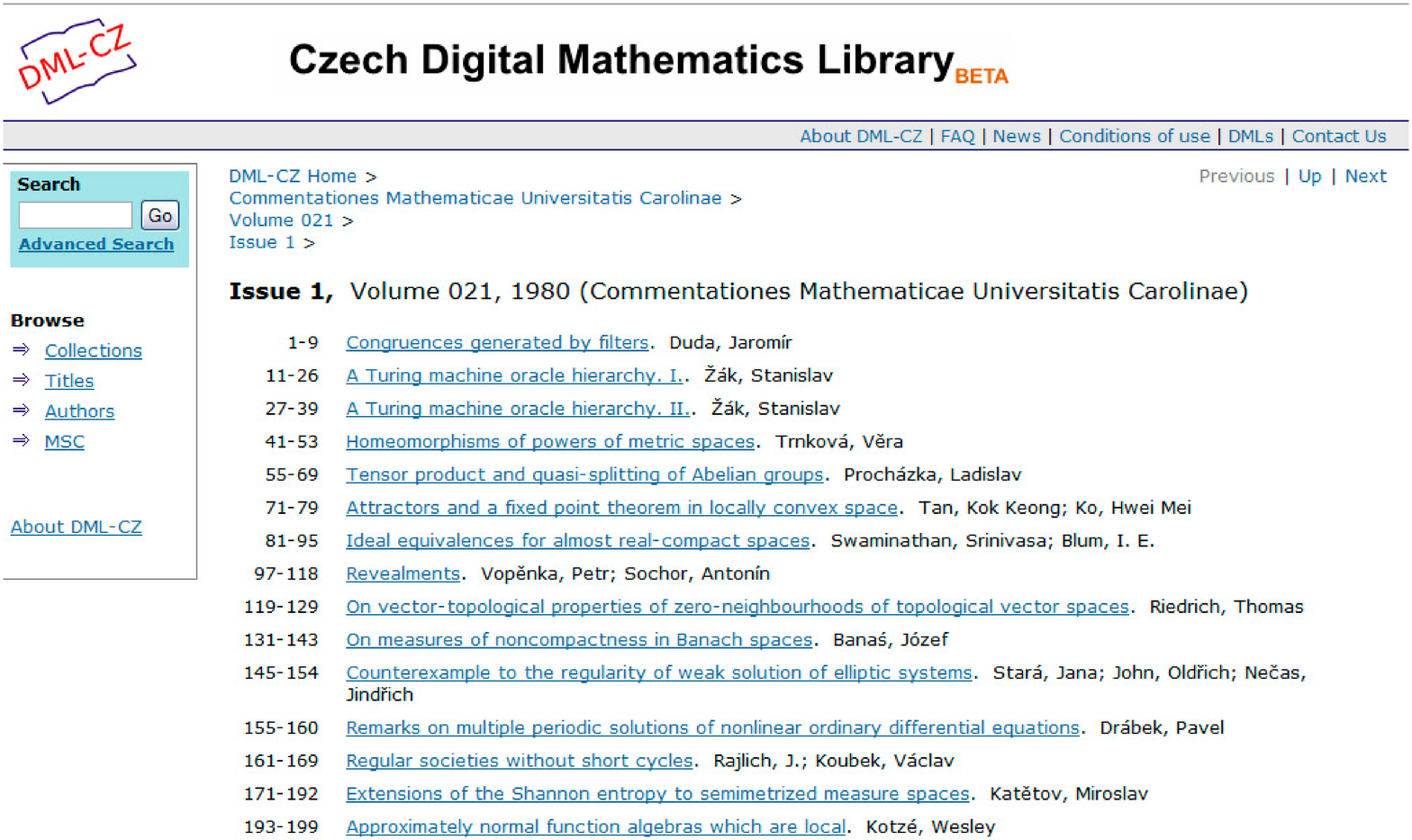}}
  \caption{The same issue of \emph{Commentationes Mathematicae
      Universitatis Carolinae}, at DigiZeitschriften (top) and DML-CZ
    (bottom).}
  \label{fig:CMUC}
\end{figure}

\begin{figure}[tp]
  \centering
  \fbox{\includegraphics[width=\zegoodwidth]{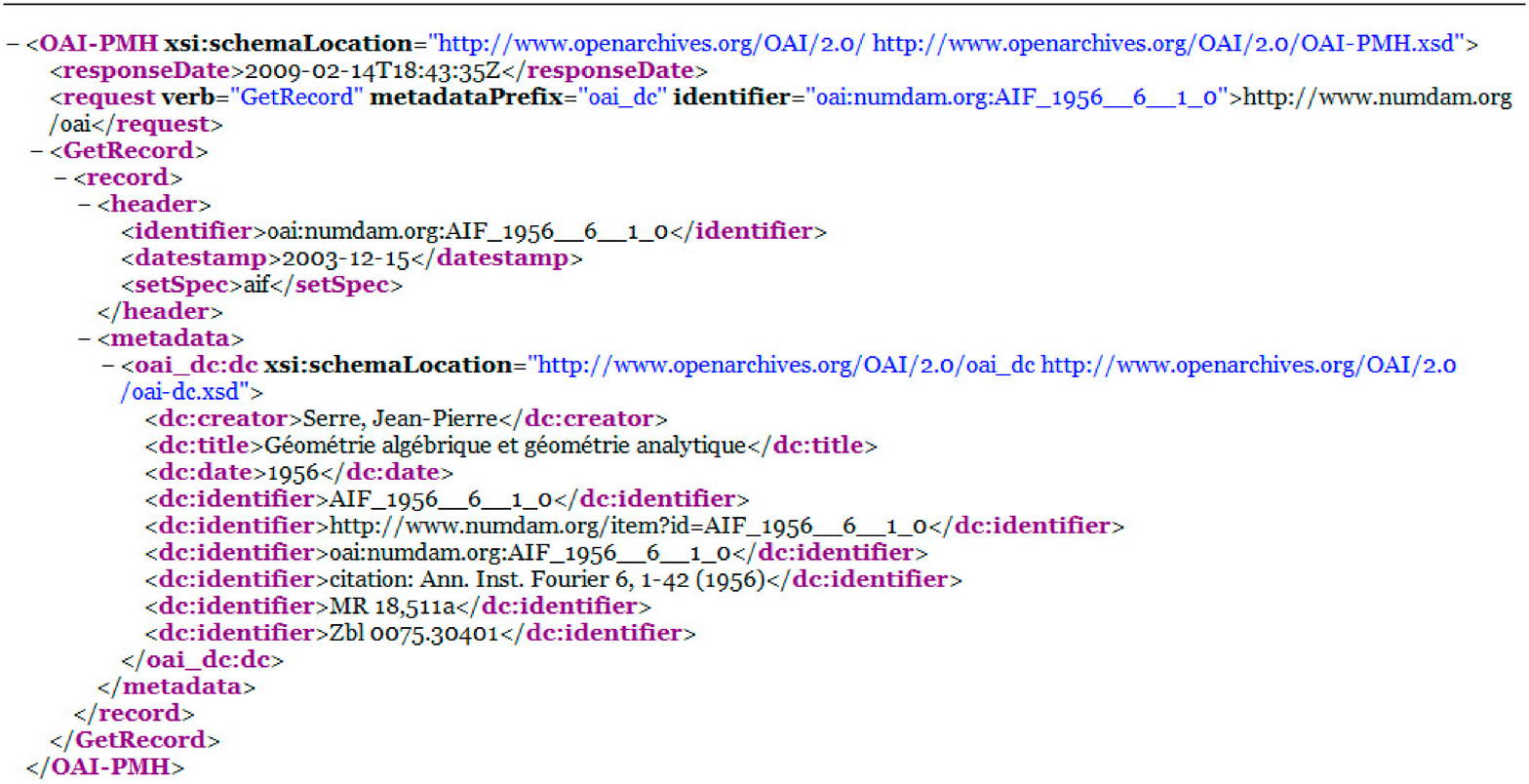}}\\[3mm]
  \fbox{\includegraphics[width=\zegoodwidth]{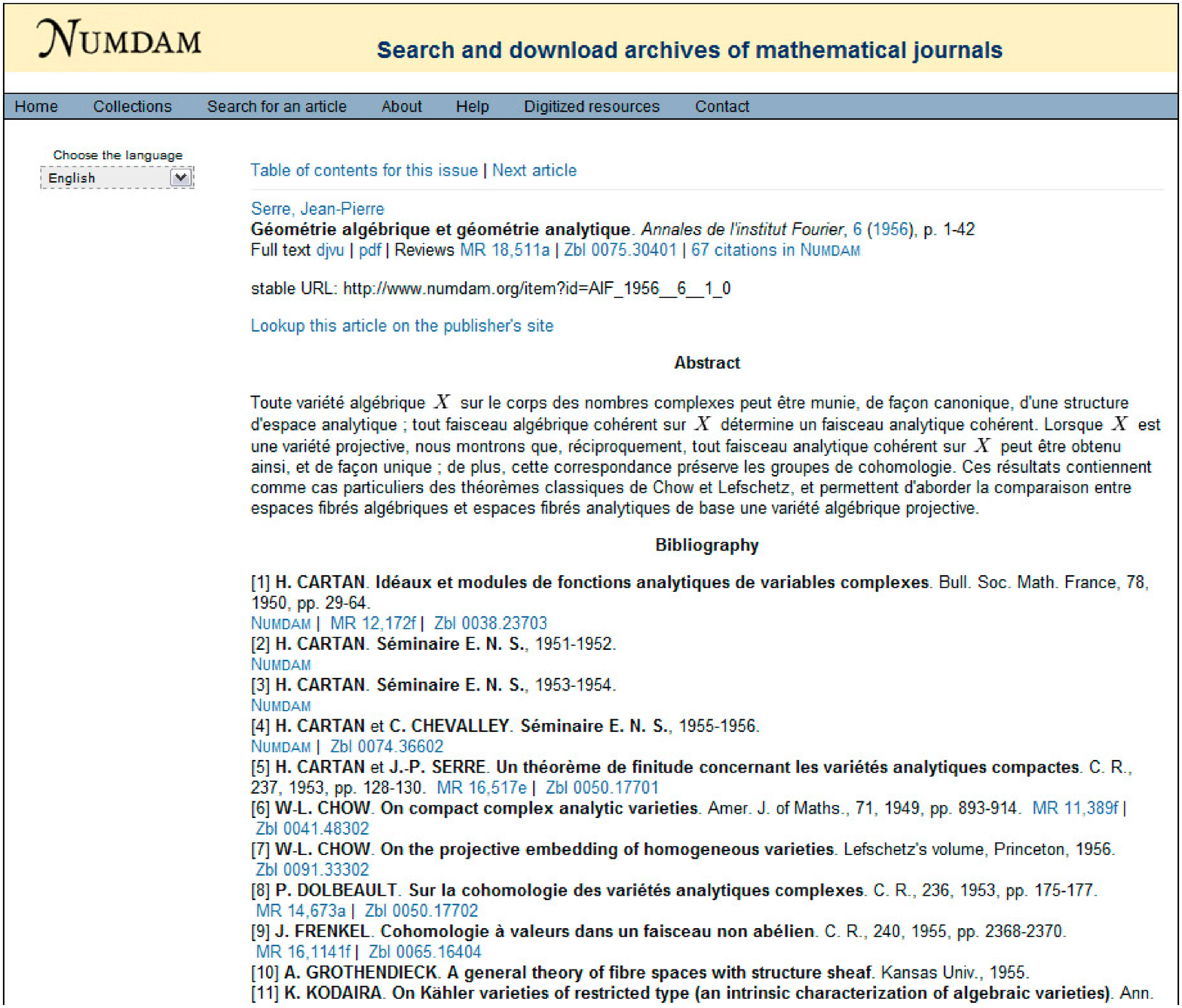}}
  \caption{The metadata of the same article of Jean-Pierre Serre at
    NUMDAM, exported through OAI-PMH (top), and exposed on the website
    (bottom).}
  \label{fig:serre}
\end{figure}

Among the already numerous institutions that care for a part of the
existing DML, no agreement has been found on metadata, although it is
the most crucial step for (inter)operability (see
figure~\ref{fig:serre} for how much metadata from the same source can
vary depending on the delivery channel). At some places, metadata is
reduced to a strict minimum, in such a fashion that it would even be
difficult to enhance it with metadata from another source (top of
figure~\ref{fig:CMUC} provides an example).

One main challenge is integrating content from a huge diversity of
providers, with very different skills and operational models. Taking
costs into account, it seems hardly possible that every single item
that is already in some sort of digital format and handled by a
reliable institution will get enriched metadata even if it is needed
to get interoperable.

This is where various MKM techniques could be called. The main point
is that items in the mathematical corpus do not live by themselves,
but inside a rich ``social network'' of similar items. 

First of all, many items are already known and referenced in some
existing catalogue or reviewed in some reviewing journal. By matching the
item in these preexisting databases, it would be possible to endow it
with more detailed metadata. Second, an item bears in itself many
meaningful links to other items that can be better known, and thus
whose metadata could be partly shared with it. If we assume that it is
possible to get some full text (OCR or text extraction) from an item,
and there to recognise author or editor contributed metadata like
keywords, MSC, bibliography, these provide links, and these links in
turn can yield some metadata. 

Thus, if you see the mathematical corpus as a graph, citations and
references as arrows, you could replace an item with insufficient
metadata into proper context by taking this into account. This could
in turn be used to complement missing metadata by metadata pulled or
synthesised from the related items.

Moreover, if one is able to capture and associate some of the
formulaes or other structured content like diagrams in an item, 
this could help create new links with items holding similar
mathematical constructs, to the same benefits.

These techniques show thus very promising ways for recovering items
with scarce metadata, taking advantage of neighbouring more fortunate
items, rather than competing with them.

\smallskip

Similarly, something has to be done regarding multilingualism, as the
legacy mathematical corpus is deeply multilingual. For recent items,
one can expect that at least some metadata does exist in English, like
title, abstract, key words, plus those derived from the MSC. For less
recent items, some metadata has been translated in English as well for
many items, either in translation journals or by English-speaking
reviewing journals. When we go farther in the past, we are stuck with
original texts spread across many languages, with no English
counterpart. Making the whole corpus searchable (with English
keywords) will require a lot of work, but some of it could possibly be
automated. That would bring back to visibility a large part of the
corpus!

Optical mathematical expression recognition coupled with formula
searching could also be a promising path to discover articles on
behalf of their scientific meaning rather than their linguistic
incarnation.

\subsection{Integration}

The biggest challenge probably sits in the area of integration.
Although so many mathematical items have already a digital version,
although a substantial proportion of these is hosted by not-for-profit
organisations willing to cooperate with international partners,
non-trivial integration of even a reduced number of these collections
has not yet happened.

Some ventures have had some success: reviewing databases turn
themselves progressively into portals offering links to the reviewed
items, which is mostly based on Crossref linking. But duplicates are
seldom handled (some Springer journals are available freely at GDZ,
while not freely at Springerlink, which owns the DOIs\ldots), and so
many small collections appear (and disappear or move) every day that
they cannot keep track of them.  What is needed is an independent
infrastructure providing the facility to register for small projects
that could upload their holdings' metadata, lookups allowing to match
databases with overlapping content, so that any bibliographical
reference can be turned into a permanent link,

In Grenoble, we maintain a small project which is meant as a
proof-of-concept for this: the mini-DML. It is an OAI-PMH harvester
with basic search interface. It has minimal requirements regarding
metadata granularity.  These minimal requirements are to attach to
each item: author(s), title, date of publication, bibliographic
reference (journal, issue, pages, etc.). It happens that, as only
simple Dublin Core is mandatory in OAI-PMH standard, and as it can be
interpreted in so many ways when describing scholarly content, we have
to develop special strategies for dealing with each repository we
harvest in order to interpret properly their article metadata in a
unified manner. Moreover, almost no field being mandatory, we gave up
dealing with many centres that would be willing to cooperate because
the metadata they publish does not meet our requirements. Highwire
Press OAI server provides author/title and that's it. Often dates
delivered are those of the metadata, or of the online posting, not of
the underlying article itself. The conclusion from this experiment is
that we cannot just wait for local libraries to deliver spontaneously
enough metadata to be integrated into larger virtual libraries, we
must give them reasons and incentives to share it, possibly using more
private communication channels.

\section{Overview of some local DMLs}

\subsection{France}
In the case of French mathematical content, we can identify many local
digital libraries already conforming to a reduced version of our
``vision''. For instance, the libraries of universities like those of
Strasbourg, Lille, or \'Ecole polytechnique, have some local content
(dissertations, lecture notes, old and rare items\ldots). The Gallica
project \cite{gallica} from the French national library has digitised
a lot of public domain books, and few mathematical journals, usually
with a 70 years moving wall with the notable exception of the
\emph{Comptes rendus de l'acad\'emie des sciences} which are there up
to 1996. This means that the CRAS, series A has found its local
dedicated institution. As Cellule MathDoc is an associated partner of
Gallica for mathematical digitisation, it should try its best to
refine the scarce Gallica metadata.

Concerning mathematical serials published in France, the picture has
dramatically changed during the recent years, as the NUMDAM programme
has succeeded beyond its initial mission. All but five currently
alive journals have agreed on digitisation of their whole backfiles,
acquisition of born-digital recent articles through export from their
publisher's platform, and open access with a moving wall of 5 years.

The four platforms that transfer their born-digital articles are
\Newitem CEDRAM: This is a MathDoc project that was set up in order to
  enable full-featured electronic edition for independent and society
  journals, based on a robust, NUMDAM compliant platform
  \cite{cedram,20bib:aveiro}. It contributes the current content of
  4~journals published by mathematics department at Bordeaux,
  Clermont-Ferrand, Grenoble, and Toulouse, one new electronic journal
  published by the French applied mathematics society (SMAI), and
  three seminar proceedings;
\Newitem Elsevier's electronic warehouse exports PDFs and XML
  metadata (header and footer of full texts) up to year 2007 for those
  three journals whose titles belong to a French academic institution
  and whose publication was outsourced with Elsevier: \emph{Annales}
  edited by \'Ecole normale sup\'erieure and Institut Henri Poincar\'e. Two
  of them changed publisher in 2008;
\Newitem EDP Sciences exports PDFs, \LaTeX\ sources, and XML metadata for
  those journals edited by the SMAI, published in the ESAIM series;
\Newitem Springer-SBM exports PDFs and XML metadata (header and footer of
  full texts) for the \emph{Publications math\'ematiques}, which are
  edited by the I.H.E.S. and distributed by Springer-Verlag,
  Berlin-Heidelberg.

In accordance with the 5 years moving wall policy, the newer material
is only present on the portal through exposed metadata, which offers
already a good deal of visibility. Using the DOI or similar persistent
URL schemes, a deep link to the article's location at publisher's site
provides access under publisher's control.

The five ``exceptional cases'' are the already mentioned CRAS, handled
by Gallica, two Elsevier journals whose titles are not currently owned
by an academic institution (which are also handled by Gallica, with a
70 years moving wall\ldots), The \emph{Journal de l'institut de
  mathématiques de Jussieu}, published by Cambridge University Press
since 2002 (which is too recent), and the \emph{Bulletin} of the
French mathematical society (SMF), whose retrodigitised version at
NUMDAM enjoys a 10 years moving wall, and no plans yet for the update
with recent articles.

\begin{figure}[tbp]
  \centering
{\includegraphics[width=\textwidth]{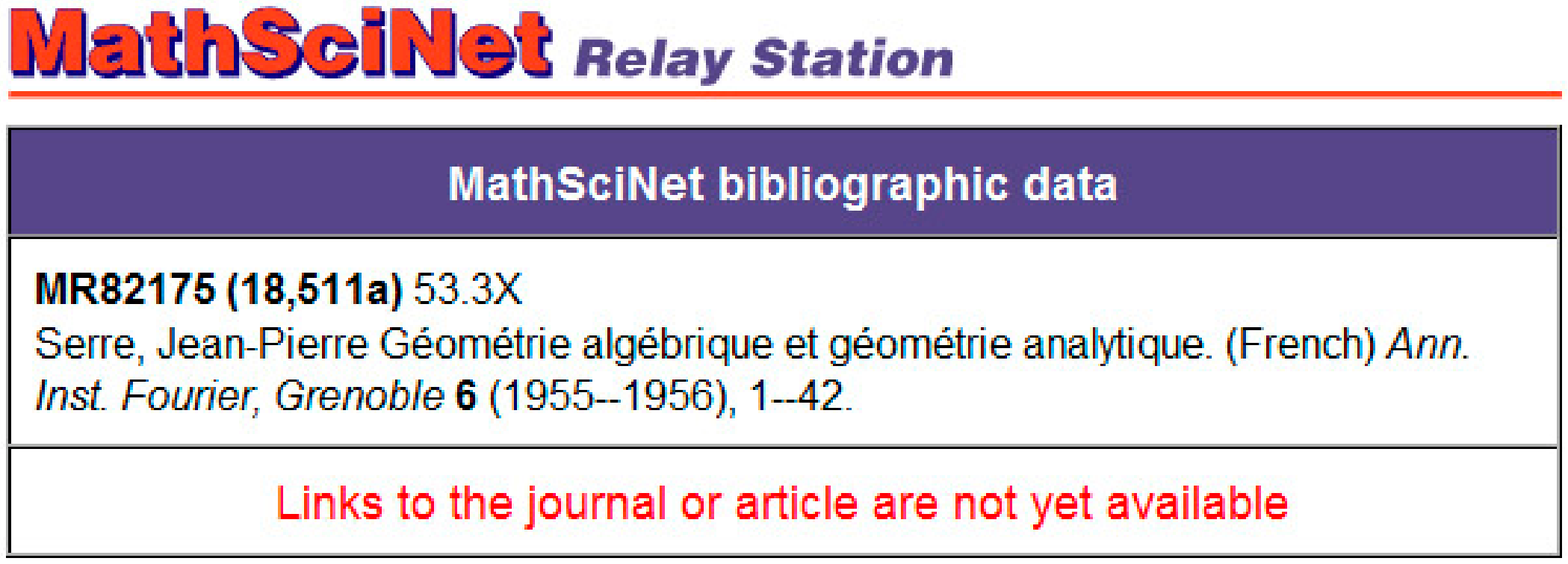}}\\
{\includegraphics[width=\textwidth]{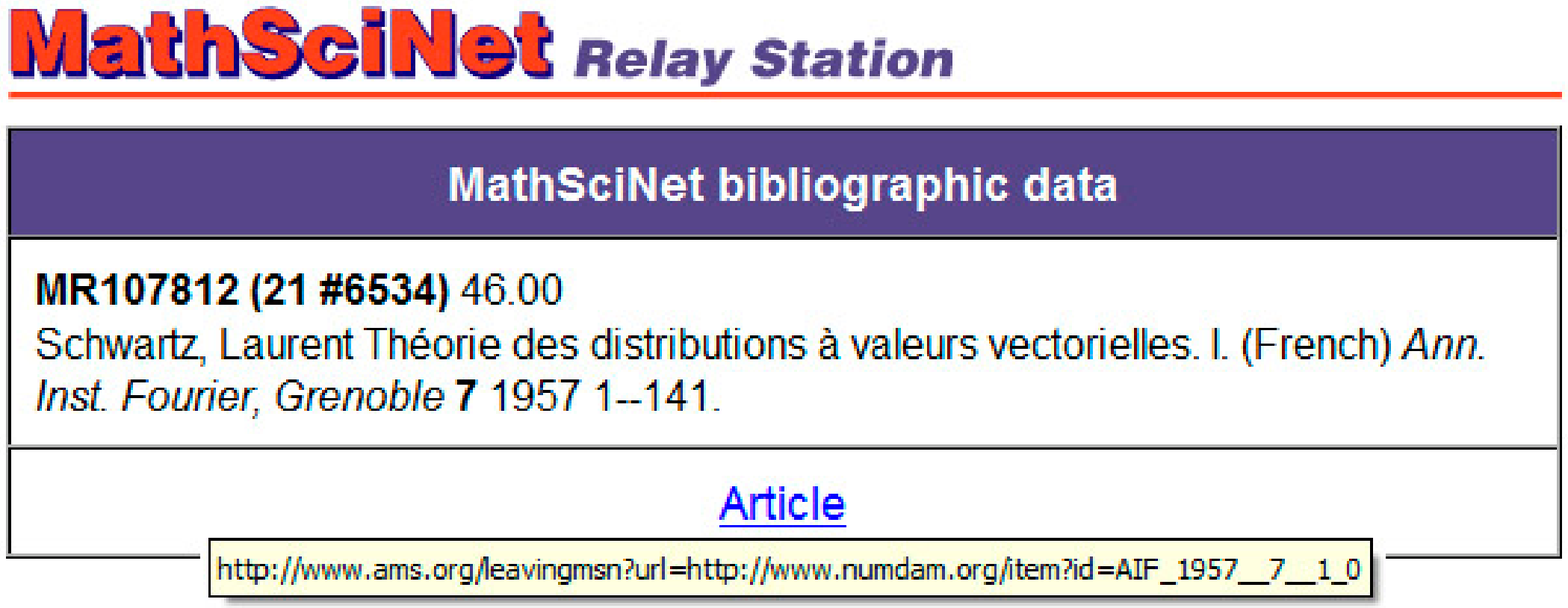}}\\[3mm]
  \fbox{\includegraphics[width=\zegoodwidth]
{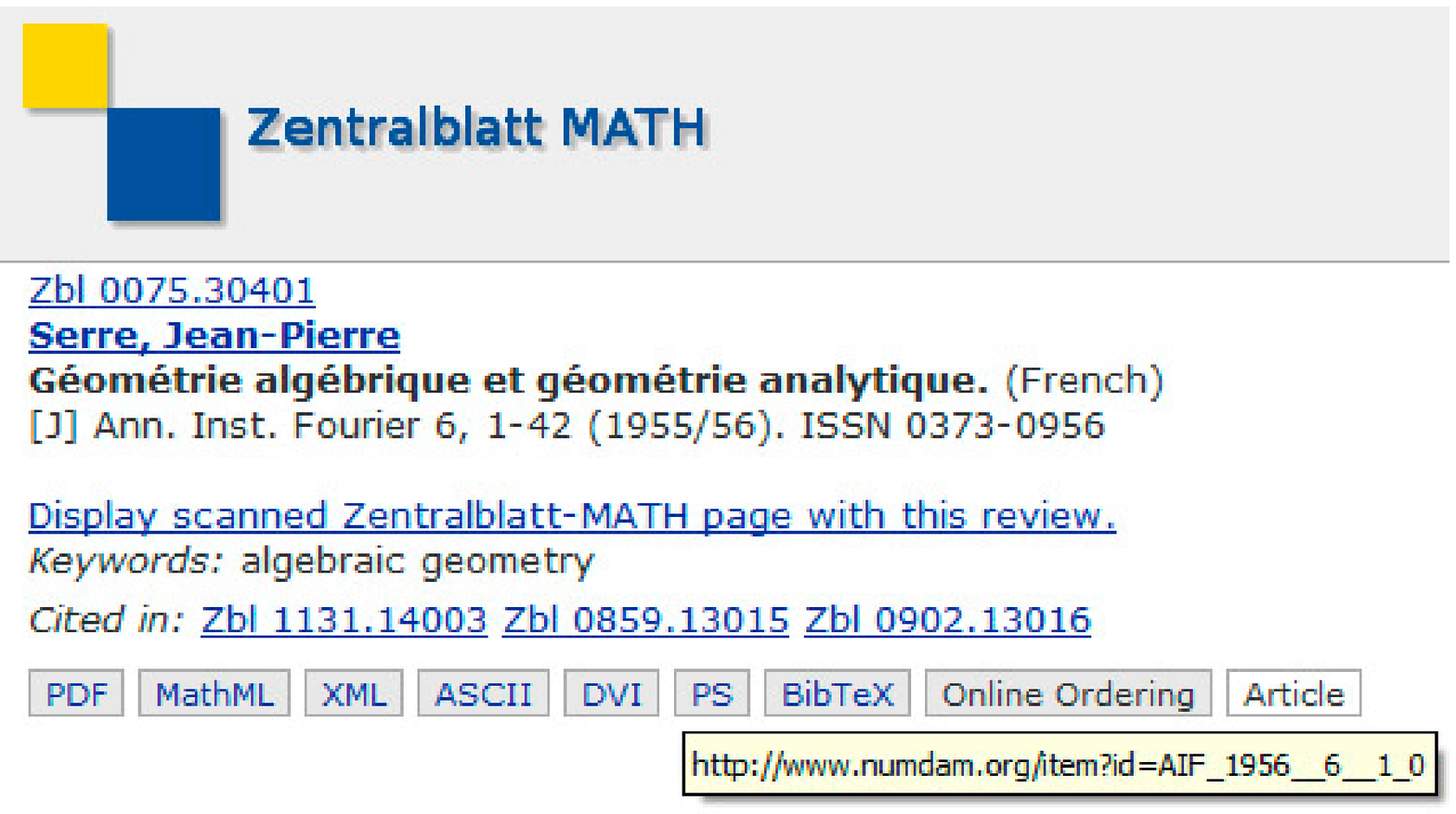}}
\caption{Two (seminal) articles from the same journal at MR with
  varying success in getting a direct link there (top) and ZM
  counterpart: there is an article link but metadata is a scanned image
  (bottom).}
  \label{fig:mrzmlinks}
\end{figure}

NUMDAM is the standard example of a good local DML, although it lacks
multilingual features (metadata of an article is derived from the
article's content, so that no English keywords are attached to
articles entirely written in French or Italian). For many users, this
might be overcome by using links to reviews in MSN or ZM, where
English metadata is often available, but this hinders wider visibility
of a substantial part of the collections. Backward links from the
reviewing databases allow the users to discover NUMDAM
articles with the sophisticated tools at their disposal there, then
access them in one click. Unfortunately, as NUMDAM doesn't currently
use DOIs, but a ``proprietary'' persistent URL scheme which is
publicised through an OAI-PMH server, many of those article links are
indeed missing (figure~\ref{fig:mrzmlinks}). 

\subsection{Czech Republic}

The DML-CZ project \cite{dmlcz} follows a similar pattern to
NUMDAM/CEDRAM. It  handles 10~retrodigitised journals (some of them
digitised in Göttingen, see figure~\ref{fig:CMUC}), some more books
and proceedings, and acquires the recent born-digital content for some
of them.

It is expected to view some intriguing new features there (search over
mathematical aware OCRed full texts, guessed MSCs and links to cosine
similar articles), but currently the metadata on which the website
operates is purely textual. However, a good point is that all articles
have at least their title translated to English. A puzzling fact to
some users is that the English title is displayed on the article's
record page rather than the original title (with the mention of the
article's language, as was standard in reviewing databases).

\subsection{North America}

Cornell's university library project Euclid \cite{euclid} is basically
an electronic publishing platform for small publishers. It has however
digitised the full backfile of some of its journals, and provides a
substantial part of its holdings as open access. As it is the primary
source for the digital items it delivers, it cannot be considered an
independent library, but as the staff running the project has
primarily a (traditional) library background, one is inclined to think
that their collections are in safe hands there. As most of the work
has to be done by publishers themselves, there is no uniform metadata
policy (bibliographical references are sometimes present, sometimes
linked, sometimes not). The fact that publishers cover all operating
costs, which they recover through subscription or fund raising,
illustrate that it is meant mainly as a service to publishers. It is
not obvious how this business model will evolve, when the costs of
maintaining the huge legacy archive of free access articles will
raise, while they present no benefit to their publishers.

JSTOR \cite{jstor} is a quite large reference library covering all
scientific disciplines. Its mathematical content is heavily biased
towards English-language serials. The growth of its mathematical
collection (by far the largest collection of its kind) during the last
decade has been mostly driven towards statistics and applications of
mathematics. This is a very well managed library, which has
unfortunately no specific features to enhance retrieval of
mathematical articles. Costs are covered through (traditional)
university libraries' subscriptions, for which it acts as a federated
digital archive service. Each subscribing library balances these costs
with savings on shelve space, which can be freed thanks to the
availability of the digitised versions to their patrons. This business
model is thus somewhat dual to that of project Euclid.

\section{Conclusion}

While the DML idea emerged a decade ago as the grand project that
would change dramatically the way we would do mathematics in the 21th
century, and was conceived as a huge, centralised---somewhat
imperialist?---process, nothing in line with these expectations has
happened. But the digital mathematical content is now omnipresent and
large.

Small scale implementation of a variation on the \mDML\ as discussed
here seems entirely feasible now. We hope that this will happen soon.
Given the high satisfaction expressed by users of isolated projects
such as those just reviewed, bridging at least two projects beyond
their current boundaries would meet high expectations in the user base
of the reference mathematical literature. 

Interconnecting most of them does not seem out of sight technically.
The main inhibiting factors are in the area of conflicting interests
among stakeholders (funding agencies want impact for the outcome of
research they support, researchers want both prestige by publishing in
selective journals and optimal dissemination and visibility of their
papers, some publishers want to generate profit from their business,
some want to secure their long term operation, libraries want to be as
open as they can afford within their budget constraints, etc.).

Apart from the French breakthrough, it seems very difficult to obtain
permission (and actual data) from commercial publishers for feeding an
independent digital library.  It also seems completely out of sight to
obtain retrodigitised backfiles from a commercial publisher who
invested in it to be sold as a special package (and succeeds pretty
wel doing so). The economic model of the \mDML\ is an entirely open
question, which depends so much upon local research and university
systems that we won't address it here. We would like to point out that
each individual DML project has found its way for making a part of the
corpus available to its patrons. Compared to this huge distributed
effort, the last step of integration seems to require marginal
overheads while it would have considerable impact.

If a bottom-up project based on the premises exposed above ever sees
the light of day, it would be in a strong position to design a
powerful environment, together with effective strategies and a
balanced policy for preserving and accessing mathematical references
over the long term.



\end{document}